\documentclass[journal]{IEEEtran}

\ifCLASSINFOpdf

\else

\fi
\usepackage{pbox}
\usepackage{multirow} 
\usepackage{booktabs}
\usepackage[keeplastbox]{flushend} 
\usepackage{makecell} % For adding second line in table
\usepackage{rotating} % For rotating the table
\usepackage{verbatim} % for Block commenting !!!
\usepackage{amssymb}
\usepackage[dvipsnames]{xcolor}
\usepackage{amsmath, amsthm, amsfonts}
\usepackage{amsfonts} % if you want blackboard bold symbols e.g. for real numbers
\usepackage{graphicx} % if you want to include jpeg or pdf pictures
\usepackage{algorithm}
\usepackage{algorithmic}
\usepackage{longtable}
\usepackage{supertabular} 
\usepackage{xcolor}
\usepackage{cuted}

\usepackage[most]{tcolorbox}
\newtcolorbox{highlighted}{colback=yellow,coltext=red,breakable}
\usepackage[update,prepend]{epstopdf}
\usepackage{cite}
\usepackage{color}
\usepackage{flushend} % for balancing last page
\usepackage{soul}
\usepackage[caption=false]{subfig}
\usepackage{etoolbox}

\makeatletter
\patchcmd{\@makecaption}
{\scshape}
{}
{}
{}
\makeatletter
\patchcmd{\@makecaption}
{\\}
{.\ }
{}
{}
\makeatother
\begin{document}

\title{A Do-It-Yourself (DIY) \\ Light-Wave Sensing and Communication Project:\\ \textit{Low-Cost, Portable, Effective,  and Fun}}
	
	\author{Sabit~Ekin,~\IEEEmembership{Member,~IEEE,}
		John~F.~O'Hara,~\IEEEmembership{Senior Member,~IEEE,}
		Emrah~Turgut,~\IEEEmembership{Member,~IEEE,}~Nicole~Colston, and
		Jeffrey~L.~Young,~\IEEEmembership{Fellow,~IEEE}% <-this % stops a space

%	\thanks{\textit{Corresponding author: Sabit Ekin}.}
	
\thanks{S.~Ekin, J.~F. O'Hara and J.~L. Young are with the School of Electrical and Computer Engineering, Oklahoma State University, Oklahoma, USA (e-mail:~sabit.ekin@okstate.edu,~oharaj@okstate.edu,~jl.young@okstate.edu).}
 
%\textit{Corresponding author: Sabit Ekin}

\thanks{E.~Turgut is with the Department of Physics, Oklahoma State University, Oklahoma, USA (e-mail:~emrah.turgut@okstate.edu).}

\thanks{N.~Colston is with the School of Teaching, Learning and Educational Sciences at College of Education, Health and Aviation, Oklahoma State University, Oklahoma, USA (e-mail:~nicole.colston@okstate.edu).}

	}

	\markboth{ARXIV VERSION UPLOADED: JULY 16 2020, EKIN ET AL.: DIY LIGHT-WAVE SENSING AND COMM. PROJ.}{}

	%\author{Hisham~Abuella,~\IEEEmembership{Student~Member,~IEEE,} Sabit~Ekin,~\IEEEmembership{Member,~IEEE,.......}
	%\thanks{H.~Abuella and S.~Ekin are with the School of Electrical and Computer Engineering, Oklahoma State University, Oklahoma, USA (e-mail:~hisham.abuella@okstate.edu,~sabit.ekin@okstate.edu).}}

\maketitle
\vspace*{-5mm}
\begin{abstract}
\textit{ Contribution:}
% \textcolor{red}{(Contributions are expected go beyond description of an approach. Briefly describe what has been learned from the study and what is distinctive and/or novel about the approach described in the study.)}
A do-it-yourself (DIY) light-wave sensing (LWS) and communication project was developed to generate interest and clarify basic electromagnetic (EM) and wireless sensing/communication concepts among students at different education levels from middle school to undergraduate.  This paper demonstrates the nature of the project and its preliminary effectiveness.
% \vspace*{1mm}

\textit{Background:}  
% \textcolor{red}{(Briefly describe the rationale for the study presented in manuscript, that is, why is the study needed? The background is expected to establish context that suggests the study has broad application in many programs across the world.)}
Concepts in wireless sensing and communication are generally considered hard to comprehend being underpinned only by theoretical coursework and occasional simulations.  Further, K-12 schools and small academic institutions may not have the resources necessary to produce tangible demonstrations for clarification.  The consequent lack of affordable hands-on experiences fails to motivate and engage students.  
% \vspace*{1mm}

\textit{Intended Outcomes:}  
% \textcolor{red}{(Briefly list desired outcomes of the approach.)}
The DIY-LWS is intended to make wireless concepts more understandable and less esoteric by linking fundamental concepts with student-relevant technologies such as solar cells, visible lights, and smartphones.  It is also intended to pique student interest by allowing them to personally assemble, operate, and explore a light-based wireless communication system.  Students complete assessments before and after the project to quantify its effectiveness. 
% \vspace*{1mm}

\textit{Application Design:}  
% \textcolor{red}{(Briefly describe the rationale for the selected instructional approach.)}
A preliminary assessment is used to determine the student base knowledge level and enthusiasm for wireless and related core topics.  Students are instructed to assemble and test their own DIY-LWS hardware to provide a hands-on experience and stimulate further exploration.  Short lectures are given to link conceptual ideas to the real-world phenomena students personally observed.  Finally, students are re-assessed to quantify any change in conceptual understanding.
% \vspace*{0.1mm}

\textit{Findings:}  
%\textcolor{red}{(Briefly summarize the findings of the study.)} 
The DIY-LWS kits have been used in multiple events with students at different levels from secondary to high schools to college. Pre- and post-assessments revealed pronounced improvements (the number of correct answers doubled) in student understanding of EM concepts. Instructors observed tremendous interest and excitement among the students during and after the experiments.
\end{abstract}

% \vspace*{5mm}


% \vspace*{5mm}

% \fi
\IEEEpeerreviewmaketitle

%\textcolor{red}{(S)abit, (J)ohn, (E)mrah, (Y)oung, (N)icole\\
%\textbf{Max 6 pages}\\
%\textbf{1.5 pages}: Abstract (S/J) + Introduction (J/Y) + Motivations, Learning Outcomes and Structure (J/Y/N) \\
%\textbf{2.5 pages}: Project Description (S/J) \\
%\textbf{1.5 page}:Assessment and Evaluation (E/N) \\
%\textbf{0.5 page}: Conclusion (E/N) + References}

%extra stuff from abstract
%project uses wireless sensing and wireless communication with visible and/or infrared light to convey basic electrical engineering and physics concepts.  The project allows students to build their own LWS system, consisting of a audio player (e.g., smart phone), light source (light-emitting diode at any color including invisible infrared), a solar panel, amplifiers, batteries and cables.

%These DIY design kits have been used in multiple events with students at different education levels from secondary through college. Pre- and post-assessments were conducted to receive feedback on their experience and evaluate their knowledge in electromagnetic spectrum, light waves, amplitude modulation, optical sensing and communication and light reflectivity. Instructors have observed a great interest and excitement during and after the experiments, and knowledge gained.

\section{Introduction}
\IEEEPARstart{I}{t} is widely recognized that early undergraduate and high-school students often have faulty conceptions about the nature of electromagnetic (EM) fields,  radiation and wireless communication \cite{bunting2009vector, kesonen2011university, plotz2016students, roussel2012difficulties, xu2011visualizing}.  These misconceptions take on myriad forms, ranging from the fundamental (e.g., the distinction between particle radiation and EM radiation) \cite{plotz2016students}, to the educational (e.g., as a subject, EM is too difficult and demanding) \cite{roussel2012difficulties}. At the practical level, students also struggle to learn how EM \emph{systems} work, since these are often presented with abstract concepts such as modulation, wave phase, noise statistics, and bandwidth.  Not surprisingly, misconceptions are also accompanied by a continuing challenge to motivate student interest in pursuing higher level education with EM or an EM-related focus.  This is despite the fact that modern society is replete with examples and applications that employ wireless communications and sensing. Misconceptions have been likewise found among primary school science educators \cite{unver2018exploration} revealing a persistence of problems in both the learning and teaching of such principles. 

It is also widely recognized that demonstrations of EM and wireless sensing/communication concepts with real-world hardware can enhance student interest and cognition \cite{bunting2009vector,sarkar2006teaching, rojas2018use}. 
The challenge here is that typical diagnostic hardware available in the high university setting is either expensive and inaccessible, or requires a significant time-investment for the instructors.  For example, some previously reported examples of EM demonstrators involve oscilloscopes and/or RF transceivers \cite{sarkar2006teaching}, or spectrum analyzers and signal generators \cite{rojas2018use}. While these are intriguing approaches, they are not cost-effective to deploy in plurality, even within a large university. At secondary schools, even one instrument is usually well outside of budgetary restrictions. In addition, a certain level of expertise and training is required to effectively utilize these equipment.    

% For example, 15\% of students (on average of two years) in electrical and computer engineering department at Oklahoma State University take advanced EM, while the most students signed up for courses in other areas such as systems, power and computer engineering.

%Here are some statistics:
%F19, ECEN 3613: 56 students
%S20, ECEN 3623: 10 Students
%F19, ECEN 4613: 8 Students
%Hence, about 15\% students take advanced EM. 
%Most students sign up for systems, power and computer engineering, but the latter is also due to the BSCpE degree
%(\hl{Are there statistics \emph{within OSU-ECE} regarding students going into EM/Photonics vs other areas? -JFO})

%The authors expertise are; wireless communication including both radio-frequency and optical light-wave communication systems, optics, photonics, engineering education, STEM education, .....
As an element of their research on EM, photonics and communications, the authors designed the do-it-yourself light-wave sensing (DIY-LWS) and communication kit with off-the-shelf parts. Each kit includes a few light sources (light-emitting diodes at any color including invisible infrared), a solar panel, two amplifiers, a speaker, batteries and cables.  The DIY-LWS kit operates as follows: 1) an audio signal is fed from an audio player (e.g., smart phone) to the input amplifier; 2) the input amplifier supplies the amplitude modulated current to the LED light, thus converting the audio signal into a light wave (\textit{transmitter}).  At low frequencies, the modulation on the light appears as a visible flickering;  3) a small solar panel receives the modulated light wave and feeds it to the output amplifier; 4) the output amplifier drives a small speaker via another audio jack (\textit{receiver}). Wireless data transmission is successfully achieved for up to a few meters of distance between transmitter and receiver.   

The kit has several pedagogical and practical strengths:
\begin{itemize}
    \item {\bf Effective:}  Students gain rapid  understanding of several EM concepts within a duration of one to two hours.
    
    \item {\bf Simple}:  The instructor's time-investment to make the kits operational is minimal (i.e., on the scale of minutes per kit).
    
    \item {\bf Affordable:}  Each kit costs only \$34.20~USD as of June~2019, thus making it possible for students to keep their kits permanently.
    
    \item {\bf Portable:}  The entire kit fits easily and neatly in a textbook-sized plastic container, as shown in Fig. \ref{fig:dyi_kit}.
    
    \item {\bf Fun:}  Instructors have observed an extremely positive student response to the kit, with strong enthusiasm to take it home and show others.
    
    \item {\bf Visual and Tangible:}  Students can both see the effects of modulation (for low frequency sound) on light and also correlate it with what they hear.  They can easily manipulate communication links for further exploration (e.g., pointing or blocking light, daisy-chaining two kits, etc.).
    
    \item {\bf Engaging:}  Students build the system themselves.  Many also reconfigure the system during their exploration.
    
    \item {\bf Accessible:}  Since the kit is largely light-based, students need not understand EM jargon such as ``microwave,'' ``coding,''  ``modulation,'' ``bandwidth,'' even though these principles can be easily taught.
    
    \item {\bf Compatible:}  The kit interfaces easily to any modern smart phone.
\end{itemize}

%Many methods of instruction fail to be engagign or comprehensible.  Theoretical and simulation are helpful but don't easily grasp the attention of students who aren't already highly motivated (or initiated) in this direction.  RF circuits are difficult to explain without sufficient background.  They are difficult to build.   Most of the functional part of the system must be pre-built or is buried in a chip.   

%This paper introduces a low-cost wireless.  Optical communication was considered Bell's most proud achievement.  

%Shows fundamental equivalency of various EM waves (Frequency difference only).  Shows how some colors may be more suitable than others for certain applications (remote control).  

%What about our preliminary and post-exercise talks?  We can introduce topics like UV damage - Night vision - familiar issues that deal with EM. %Goal:  to provide a platform for explaining numerous fundamental truths about EM radiation and in particular its use in wireless tech.

This paper describes in detail the DIY-LWS hardware, project instructions and experiments, and our observed pre- and post-assessment data (to date).  The data corroborates the conclusion that the project improves student comprehension and increase their enthusiasm and interest in EM and EM-related concepts, particularly wireless communication. 

These DIY-LWS kits were introduced to high school and incoming freshmen students during numerous outreach programs and recruitment events at Oklahoma State University such as National Lab Day, Discovery Days, Summer Bridge programs, and high school student visits. Overall students' response to the DIY-LWS kit was fantastic. ``Wow! But how?'', ``Mind-blowing'', and ``This is black magic. How does this work?'' are just three of the reactions that we received during the outreach events that is showing their enthusiasms. These comments describe how our effort triggered the curiosity of students and their teachers, which is a powerful motivation for future research and development. 

The ensuing discussion is structured as follows. In Section \ref{leaning}, learning outcomes are listed. The project description including the transmitter and receiver structures are presented in Section \ref{projdesc}. Assessment and evaluation results are given in Section \ref{assesement}. Finally, concluding remarks are drawn in Section \ref{conclusion}.

\section{Learning Outcomes}\label{leaning}
At the end of the DIY-LWS project, students should be able to accurately:

\begin{itemize}
    \item Describe how amplitude modulation conveys data.
    
    \item Describe how optical light can be used for communication and/or wireless sensing.
    
    \item Describe signal amplification.
    
    \item Differentiate EM waves on different parts of the spectrum (e.g., visible vs. invisible).
    
    \item Describe the difference between wavelength, frequency, and speed of light.
    
    \item Estimate approximate values of frequency and wavelength of visible light waves and compare those to microwaves.
    
    \item Describe signal relaying, blocking, and attenuation.
    
    \item Describe the nature and differences of light transmission, reflection and absorption.
    
    \item Understand how next-generation light-based communications (a.k.a. Li-Fi: Light-Fidelity \cite{haas2015lifi}) will work.
\end{itemize}
 
%Prior to start assembling the DIY kits, students were given a 15-minute presentation covering the basics of electromagnetic spectrum including optical spectrum, wavelength \& frequency \& speed of light relation, amplitude modulation and light-wave communication systems such free-space optical communication (a.k.a. laser communication), infrared communication as in remote controllers, visible light communication (a.k.a. Li-Fi), and fiber-optic wired communication.   

%Notable advantages:  Students can see the effects of modulation (for low frequency sound) on the light and correlate it to what they hear.  There is no need to use jargon including Wi-Fi, gigahertz, megahertz, modulation/demodulation, coding.  Students can use their home remote control or smart device with no special app.  Students are  familiar with most of the elements (lights, solar cells, smart phones, speakers).  Despite there being no scale production, the cost is still less than \$30 for the required parts.

\section{Project Description}\label{projdesc}
\begin{figure}[t]
\centering
\includegraphics[width=0.42\textwidth]{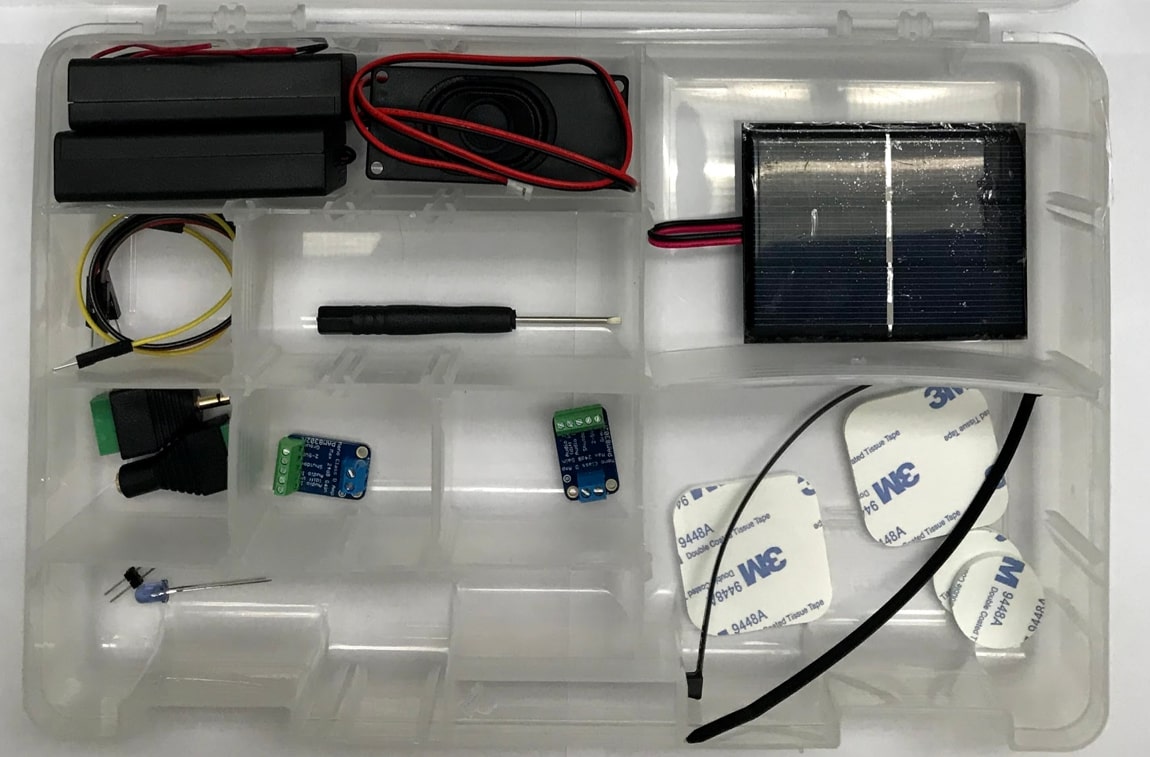}
\caption{DIY kit for light-wave sensing and communication.}
\label{fig:dyi_kit}
\end{figure}

In this project, the students integrate and test their light-wave sensing and communication projects by using parts provided in the DIY kits (See Fig. \ref{fig:dyi_kit}). The \textit{only tool needed} to put together the DIY kit is a 2.0mm flat head screwdriver (included in the kit). 

%Recall that the working principles are: 1) the \textit{transmitter} takes audio signal from an audio player device (e.g., smart phone) via audio jack, 2) the audio signal varies supplied power level for LED light (analog amplitude modulation), hence flickering occurs and audio signal transmitted through light, 3) the \textit{receiver} senses these small variations of light by a using low-cost solar panel, then sends to the speaker via another audio jack. 

By using the parts in their kits, the students first construct the transmitter and receiver, then subsequently test their light-wave sensing communication projects by playing different audio files from their own devices. They also experiment with differently colored LEDs, including ultraviolet, blue, green, yellow, red and invisible infrared and compare their performance. Additional experiments include using a remote controller (uses infrared light), double amplification by two amplifiers, relaying by using cascading two transmitter units,  blocking the line-of sight optical link with paper, and using mirrors to convey the wireless signal via a reflected path.

For completeness, and to allow the reader to reproduce our project and the results, we have provided all the details for the project including the instructions given to students. 

\subsection{Transmitter}
\begin{figure}[t]
\centering
\includegraphics[width=0.44\textwidth]{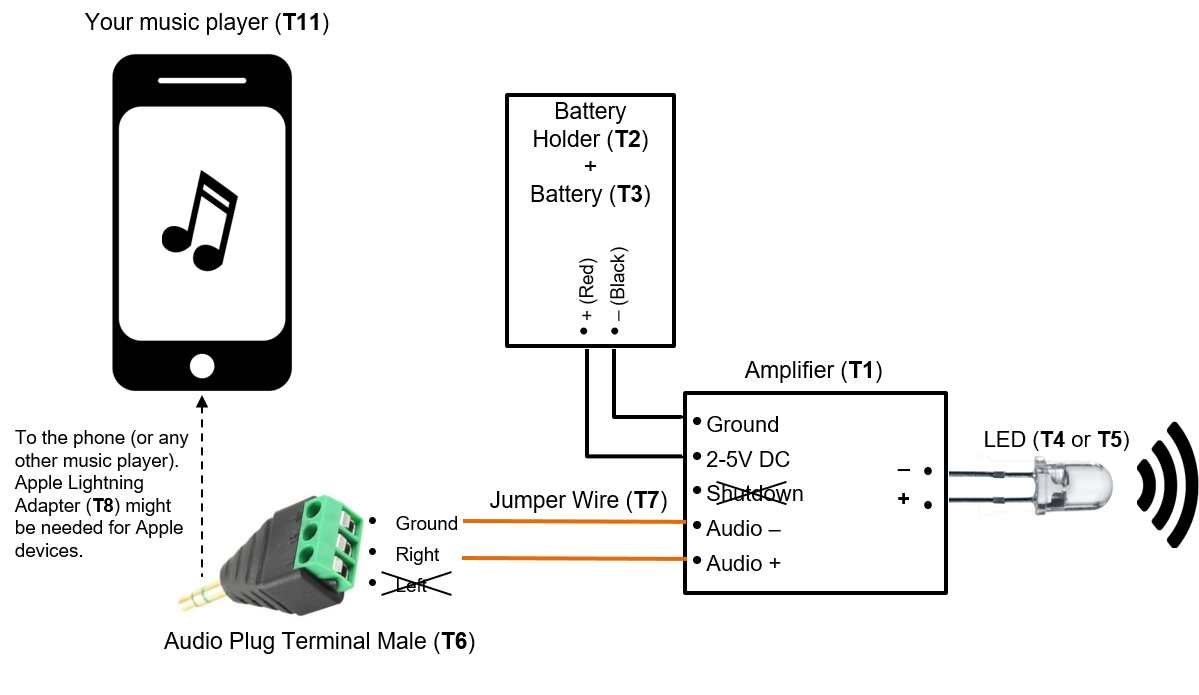}
\caption{Transmitter block diagram.}
\label{fig:tx_block_diag}
\end{figure}

The list of components needed for building the transmitter is given in Table \ref{tab:tx_items}. 
The primary components are the amplifier, LED, battery and audio player (e.g., a music player). The rest of the parts are needed for connections and sturdy packaging such as jumper wires, zip ties and adhesive foam tape. 
Fig. \ref{fig:tx_block_diag} shows transmitter block diagram, in which the audio signal is taken from a music player (e.g., a smart phone) via audio jack and sent to the amplifier. The LED light (of any color) is connected to the output of the amplifier to transmit audio signal. 

\begin{table}
    \centering
   \caption{List of components for transmitter.}
\includegraphics[width=0.48\textwidth]{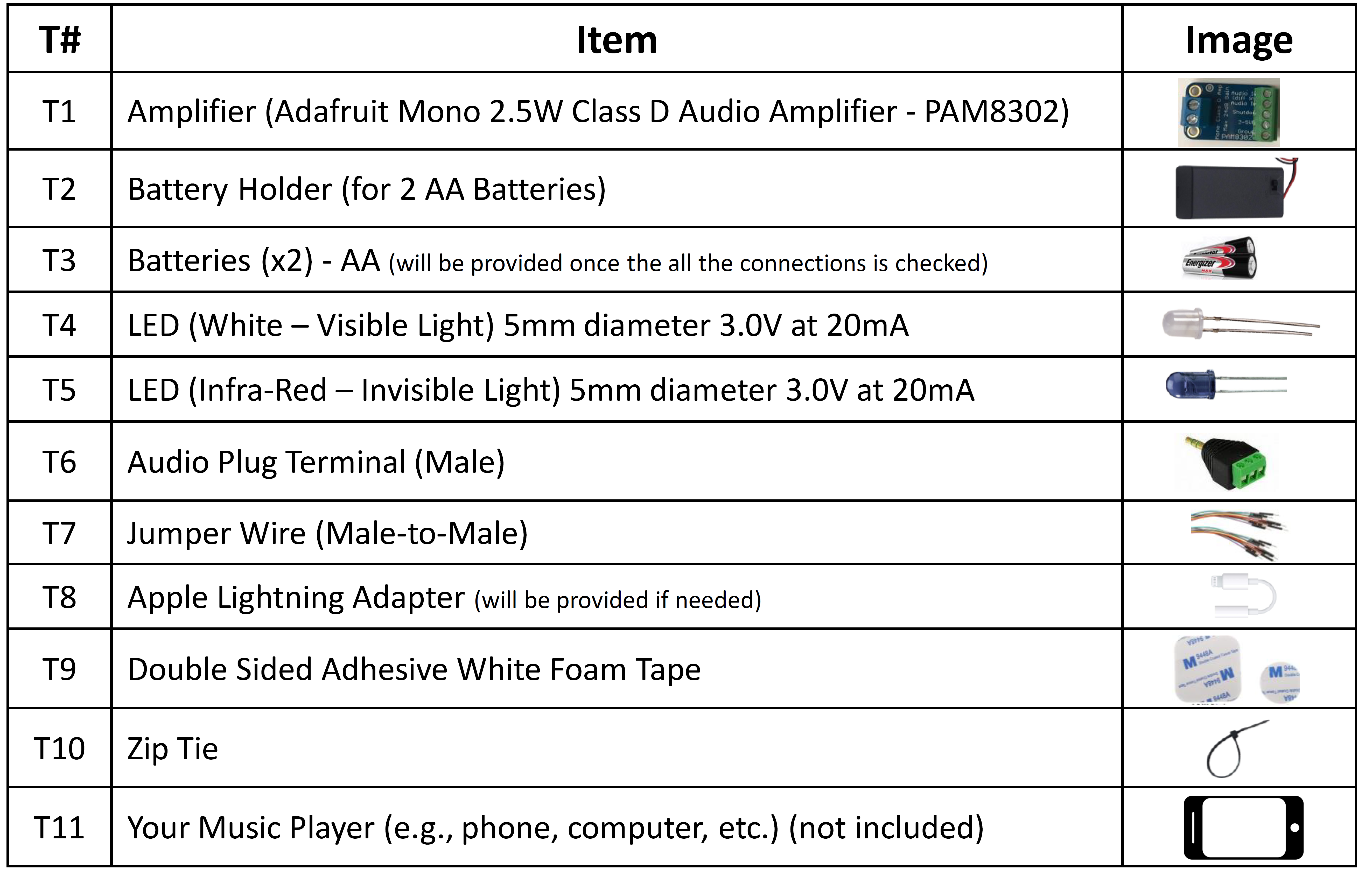}
    \label{tab:tx_items}
\end{table}

\begin{figure}[t]
\centering
\includegraphics[width=0.44\textwidth]{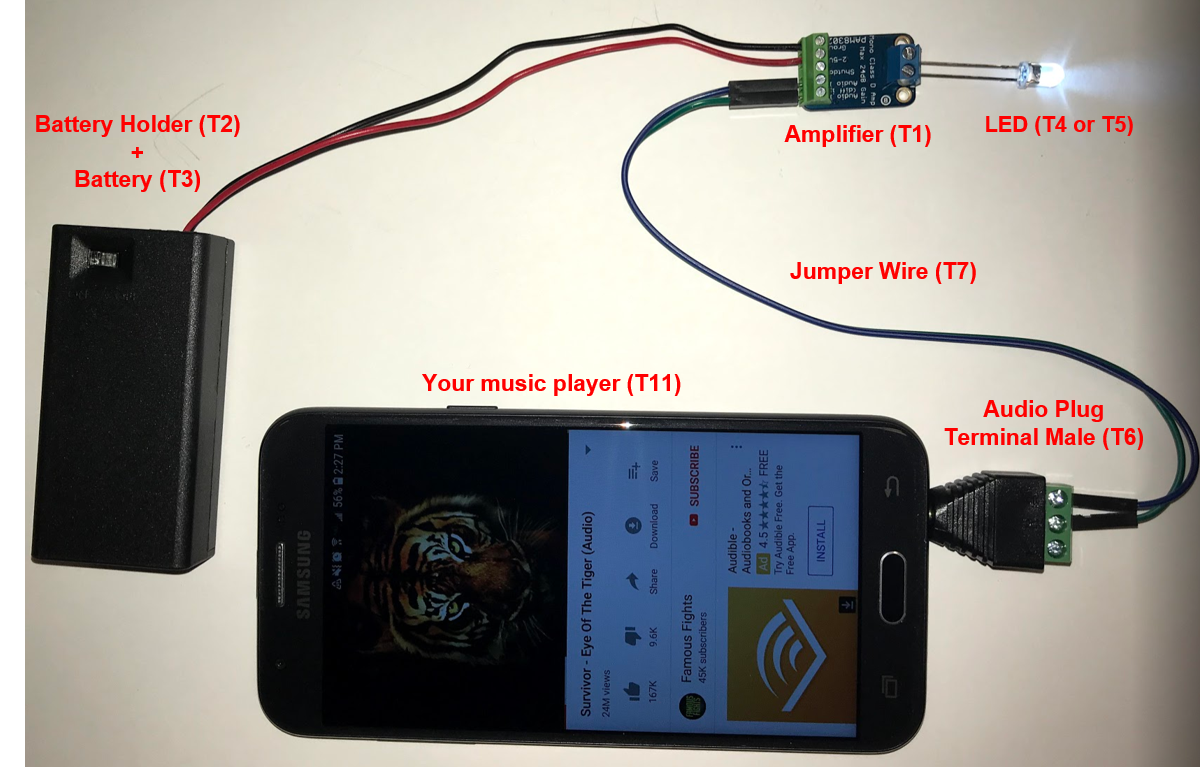}
\caption{Assembled transmitter.}
\label{fig:tx_real_setup}
\end{figure}

The following instructions were given to students for assembling their transmitters: 
\begin{enumerate}
    \item Connect Jumper Wire (T7) to Audio Plug Terminal Male (T6).
    \item Connect the other end of the Jumper Wire (T7) to Audio Amplifier (T1).
    \item Connect Battery Holder (T2) to Amplifier (T1). Polarity must be correct!
    \item Connect White LED (T4) to Amplifier (T1).
    \item Connect Audio Plug Terminal Male (T6) to your music player (e.g., phone).
    \item Your transmitter setup is complete. You can now start putting your receiver together. 
\end{enumerate}
Fig. \ref{fig:tx_real_setup} shows a completed transmitter setup after following these instructions. 

\subsection{Receiver}
\begin{figure}[t]
\centering
\includegraphics[width=0.48\textwidth]{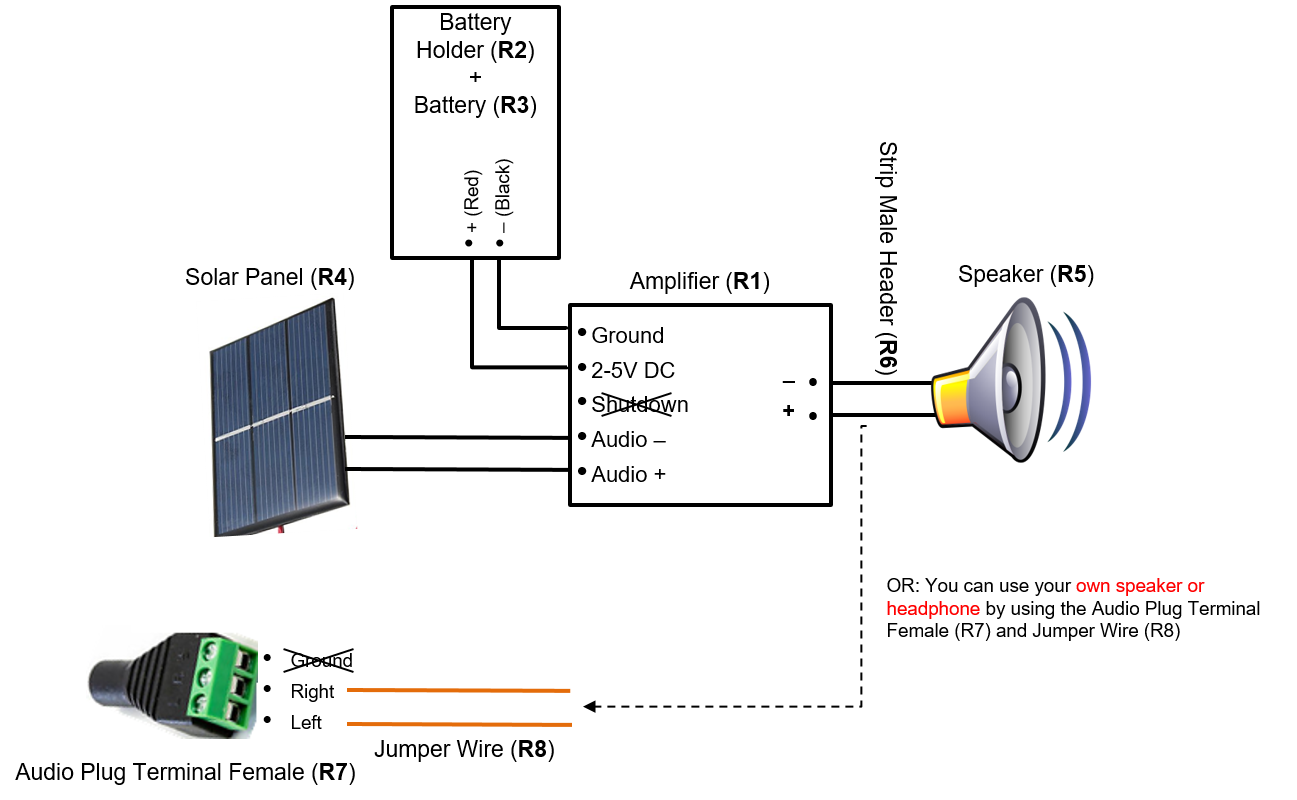}
\caption{Receiver block diagram.}
\label{fig:rx_block_diag}
\end{figure}

The list of components needed for building the receiver is given in Table \ref{tab:rx_items}. 
The primary components are the amplifier, solar panel, battery and speaker. Similarly, the rest of the parts are needed for connections and sturdy packaging such as jumper wires, strip header, zip ties and adhesive foam tape.
Fig. \ref{fig:rx_block_diag} shows the receiver block diagram, in which the solar panel detects tiny variations of LED light (due to the audio signal) and sends them to the amplifier. The amplifier outputs this signal to a small speaker provided in the DIY kit, personal speaker or headphone by using a female audio plug.

\begin{table}[t]
    \centering
   \caption{List of components for receiver.}
\includegraphics[width=0.48\textwidth]{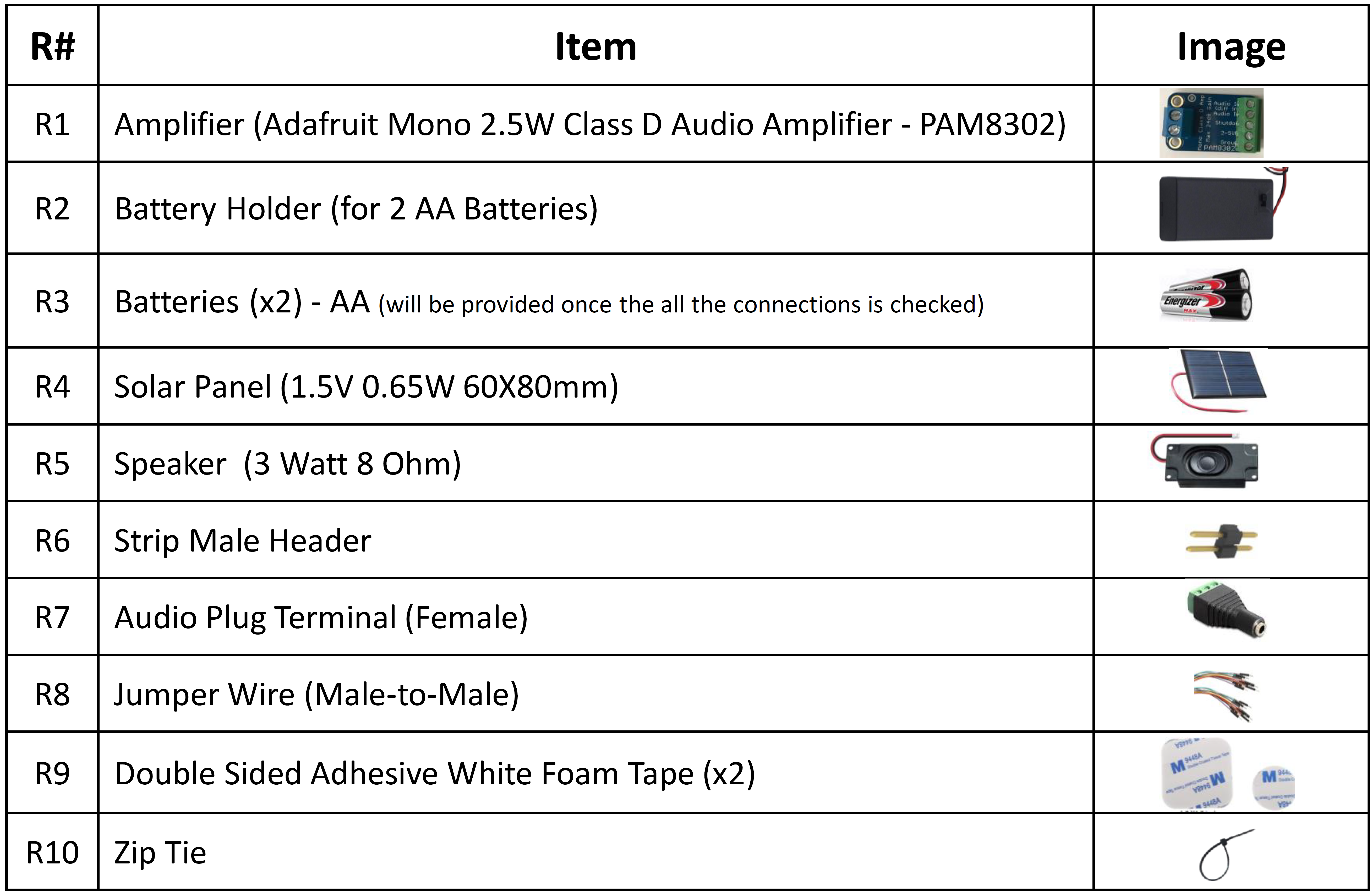}
    \label{tab:rx_items}
\end{table}

The following instructions were given to students for assembling their receivers:
\begin{enumerate}
    \item Connect Battery Holder (R2) to Amplifier (R1). Polarity must be correct.
    \item Connect Solar Panel (R4) to Amplifier (R1). 
    \item Peel off plastic cover from Solar Panel (R4).
    \item Connect Speaker (R5) to Amplifier (R1) by using Strip Male Header (R6). 
    \item Your receiver is complete. You can now start testing your system and experiment with different scenarios. 
\end{enumerate}

\begin{figure}[t]
\centering
\includegraphics[width=0.4\textwidth]{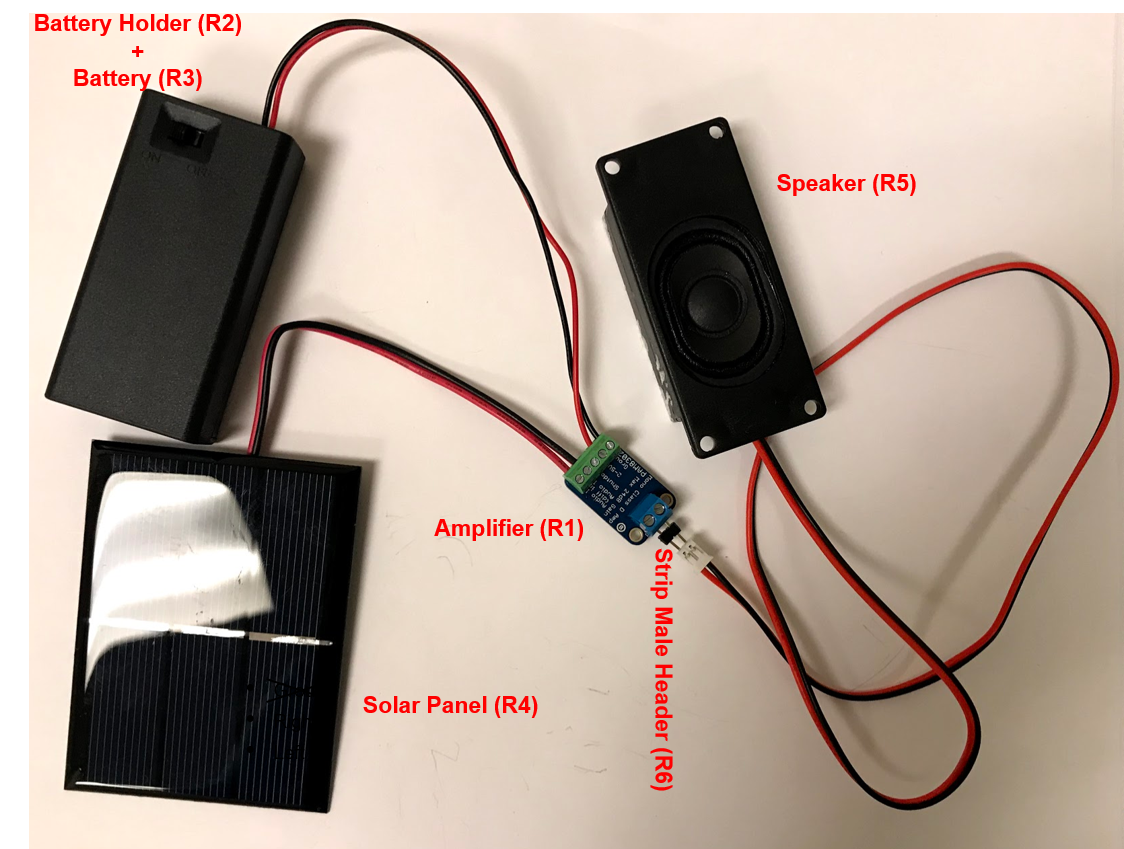}
\caption{Assembled receiver with included speaker.}
\label{fig:rx_real_setup}
\end{figure}

\begin{figure}[t]
\centering
\includegraphics[width=0.42\textwidth]{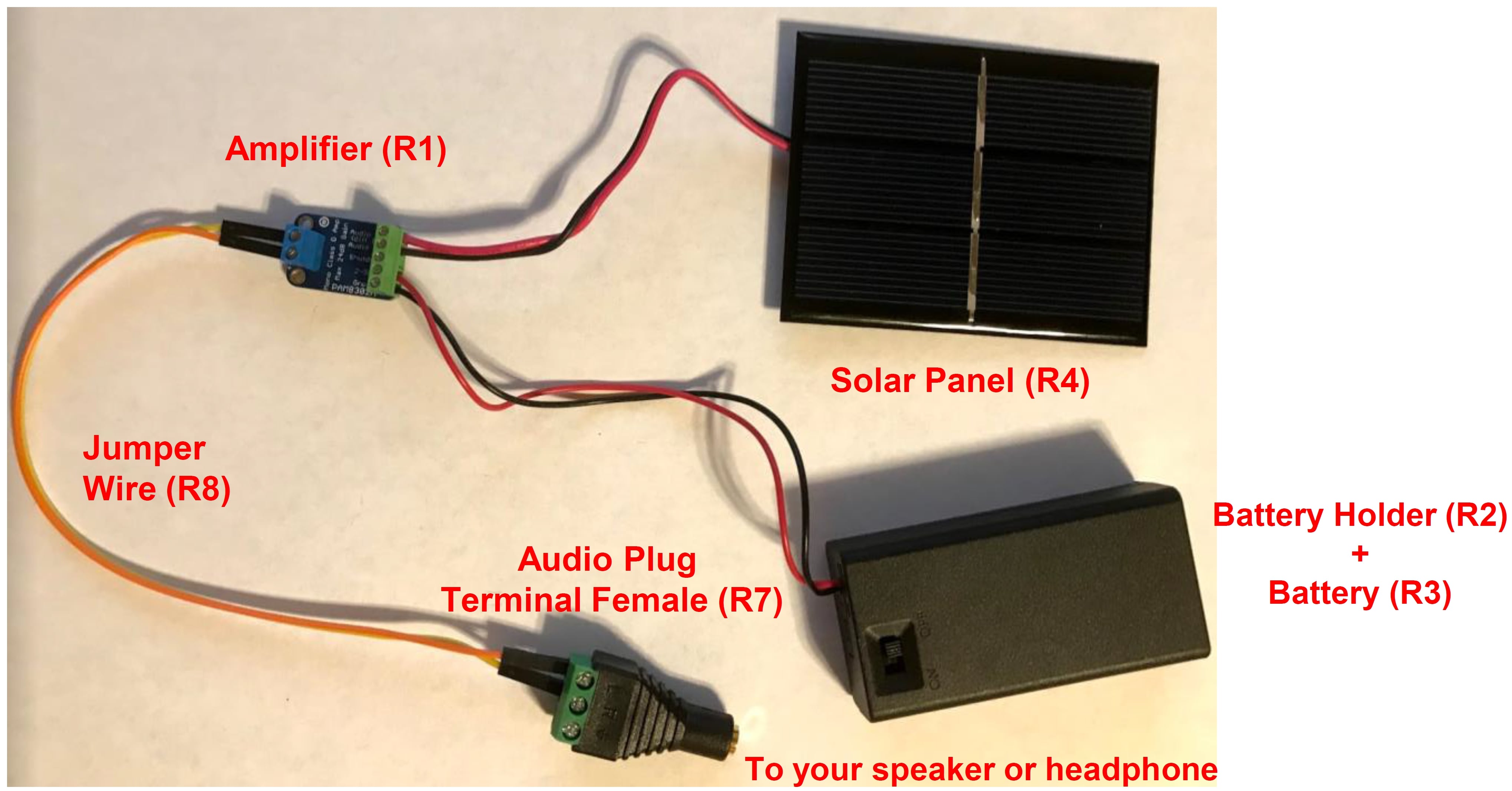}
\caption{Assembled receiver ready to be connected any speaker/headphone.}
\label{fig:rx_real_setup_own}
\end{figure}

Fig. \ref{fig:rx_real_setup} shows a completed receiver setup after following the aforementioned instructions. 
The speaker provided is small and low-power. Students can use their own speaker or headphones by performing the following steps (See Fig. \ref{fig:rx_real_setup_own}):
\begin{enumerate}
    \item Disconnect Speaker (R5) from Amplifier (R1). 
    \item Connect Jumper Wire (R7) to Audio Plug Terminal Female (R7).
    \item Connect the other end of Jumper Wire (R7) to Amplifier (R1).
    \item Plug the Female Audio Plug Terminal (R7) to your own speaker or headphone.
\end{enumerate}

\subsection{Testing and Experiments}
\begin{figure}[t]
\centering
\includegraphics[width=0.4\textwidth]{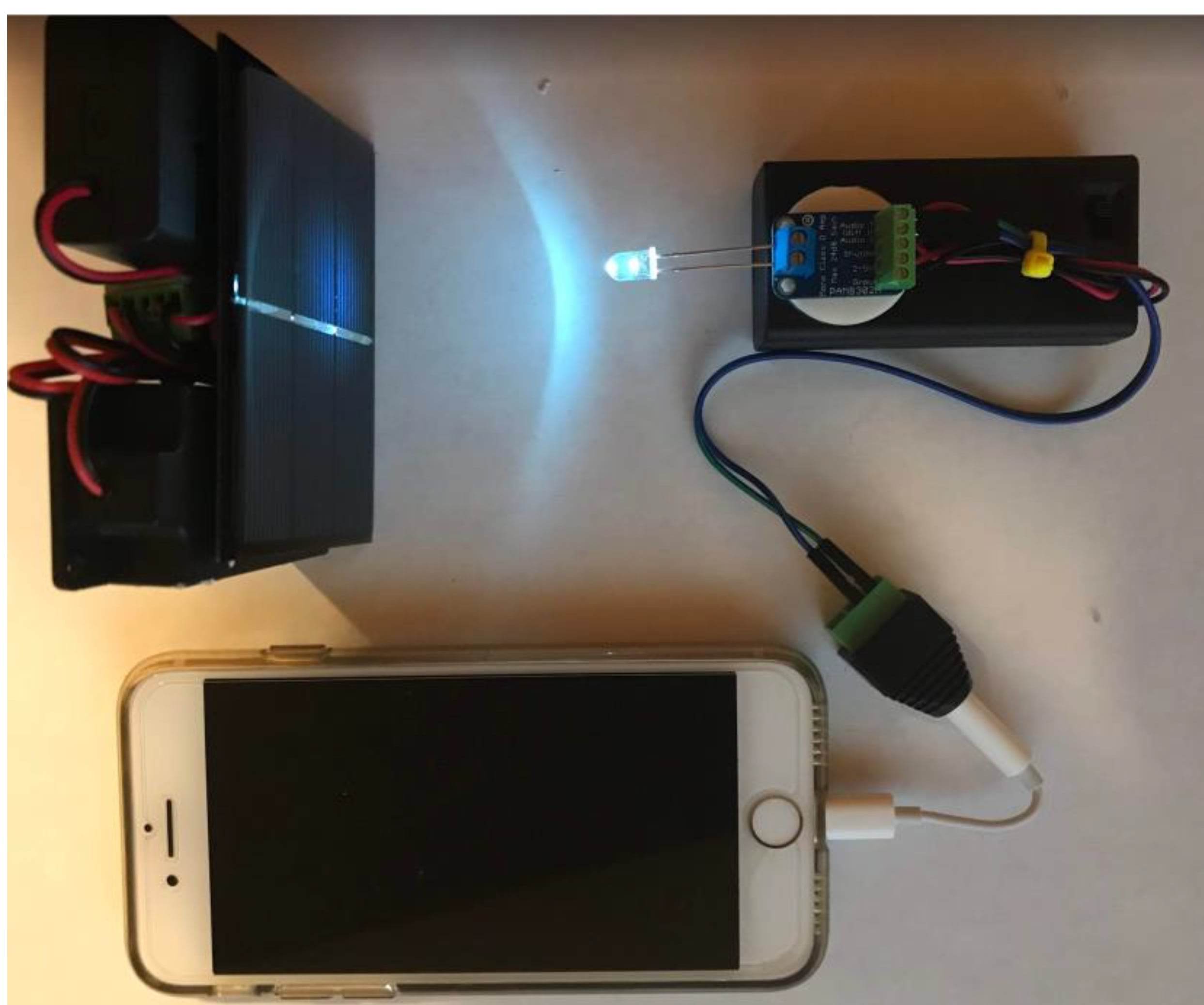}
\caption{Completed light-wave sensing and communication setup.}
\label{fig:final_touhces}
\end{figure}

The completed setup with both transmitter and receiver is provided in Fig. \ref{fig:final_touhces}, in which zip ties and adhesive foam tape was used to make an attractive and sturdy setup. 
Assembling the whole setup should take about 15 to 20 minutes on average. In our workshops, students spent approximately two hours to test and experiment with their systems. 

In order for the system to work at its maximum performance, students were recommended to 1) set the adjustable amplifiers to maximum gain, 2) peel off the plastic cover from the solar panel (if there is any), and 3) test in an indoor environment, as the sunlight can impact the performance, i.e., high noise level. 
Observed common mistakes were: 1) wrong battery connection which could damage amplifiers, 2) not turning on the power switches on battery holders for both transmitter and receiver, 3) not pointing LED (transmitters) to the solar panel (receiver), and 4) not playing an audio file at an adequate volume level. 

Once students had a working setup, they were instructed to perform the following experiments and answer corresponding questions.

\begin{figure}[t]
\centering
\includegraphics[width=0.42\textwidth]{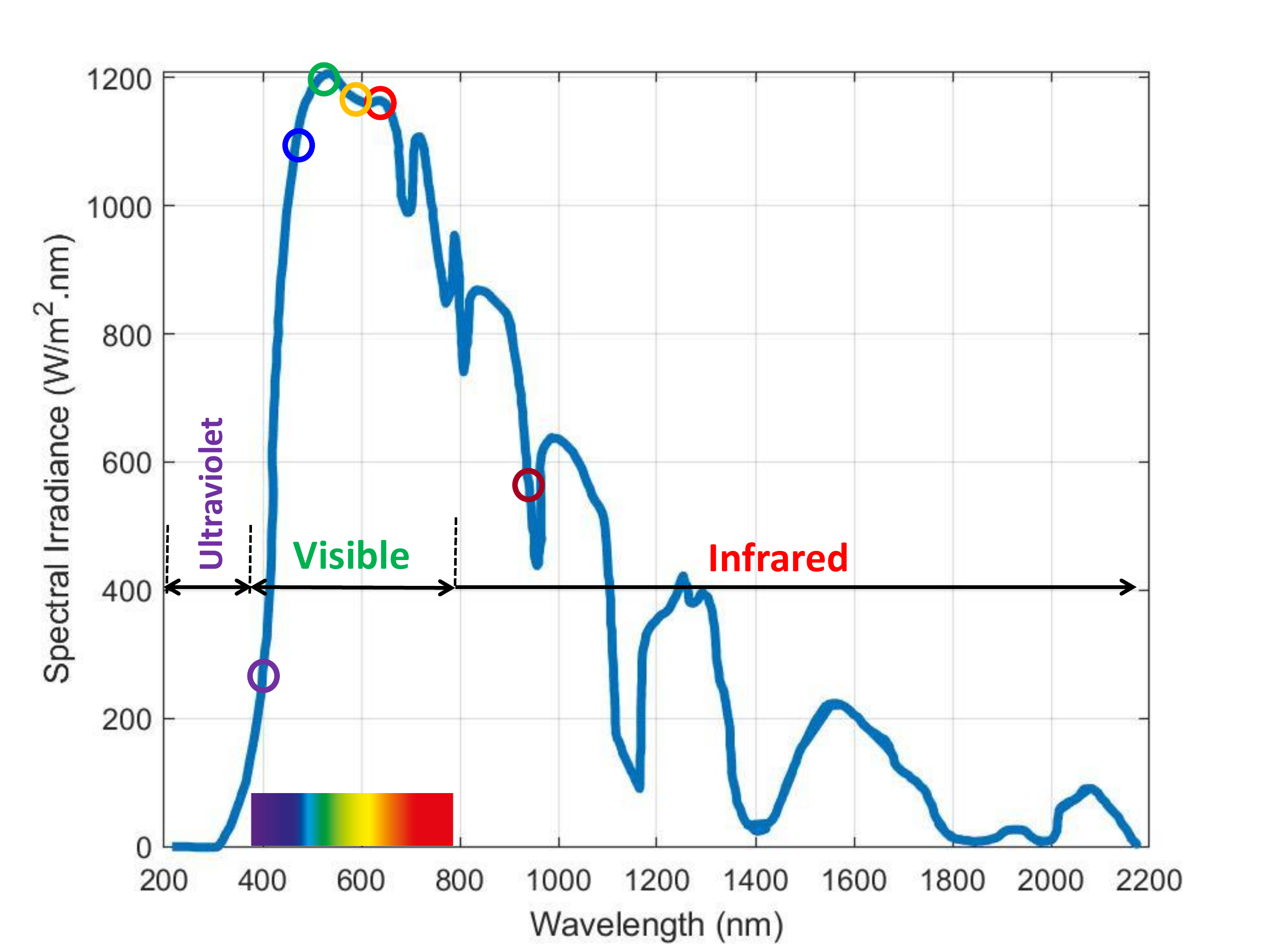}
\caption{Solar panel spectral irradiance. Colored circles shows experimented LED for this project; ultraviolet (400 nm), blue (465 nm), green (520 nm), yellow (590 nm), red (640 nm), infrared (940 nm) (regenerated from \cite{solar_panel}.)}
\label{fig:solar_panel_spectrum_diag}
\end{figure}

\begin{enumerate}
    \item Use different colors of LED light: white, ultraviolet, blue, green, yellow, red, (invisible) infrared (See Fig. \ref{fig:solar_panel_spectrum_diag} for tested light wavelengths). Does it still work? How does it compare in terms of distance and audio volume level?
    \item Use (provided) TV remote. Press different buttons of the remote while pointing it towards the solar panel. What do you hear?
    \item Place a piece of copy paper between the LED light and solar panel. Can you still hear the transmitted audio? 
    \item Use (provided) mirror to experiment with how reflected light can still transmit data. Does it still work? 
    \item Test your system under sunlight (outside or next to a window). Does it still work? 
\end{enumerate}

Our observations to the questions above are: 
\begin{enumerate}
    \item The setup worked for all colors of LED including invisible infrared and ultraviolet. The lowest performance was observed when using ultraviolet, and highest performance from infrared, which is attributed to narrow beamwidth of infrared LED and low impact from ambient light. Different distances were tested. The wireless transmission distance achieved was around 3m to 5m with infrared and other colors achieved 0.5m to 3m depending on ambient light level. 
    \item When pressing TV remote while holding toward the receiver (solar) panel, students were able to hear on-off pulses, which is the well-known space-coding method in which the length of the spaces between pulses of light represent a one or a zero.
    \item When a piece of copy paper is placed between the LED light and solar panel, students we able to still hear but with lower volume. 
    \item Students were able to send data via reflected light by using a mirror.
    \item Sunlight impacts the system with high noise level, however the student were still able to hear.  
\end{enumerate}

Due to their low-cost, we gave these kits to students at no-cost. This gave them extra freedom 
to experiment with different scenarios and to demonstrate their kit to families, friends and teachers. In addition to the aforementioned experiments, we let students use their imagination and create their own experiments. A few of the experiments that students discovered and successfully tested were: 1) using two amplifiers in parallel to increase the range, 2) ``relaying'' by using one transmitter and two receivers, the students connected an LED (instead of speaker) at first receiver's output and relayed the received signal, and 3) sensing (listening) the ambient light frequency by holding the receiver towards ambient lights. The fact that the low cost solar panel could act as a sensitive and responsive communication receiver was one of the most impressive aspects of this project, both to students and teachers. In addition, students found the ability to sense infrared light highly intriguing.

\section{Assessment and Evaluation}\label{assesement}
% \hl{(Need to specify what student groups have been evaluated, not just that "several" were - JFO)}

The DIY-LWS kits were introduced during two summer outreach experiences for high-school senior students and incoming freshman at Oklahoma State University. A total of 37 students who participated in our assessments, including 21 high school students from 18 different high schools and 16 incoming freshman from 16 different high schools. Students spent approximately 3 to 4 hours receiving instructions on assembling their kits and testing various functions. A hard-copy, pre- and post assessment survey was administered immediately prior to and after their participation. 

\subsection{Content Knowledge}
\begin{figure}[t]
\centering
\includegraphics[width=0.45\textwidth]{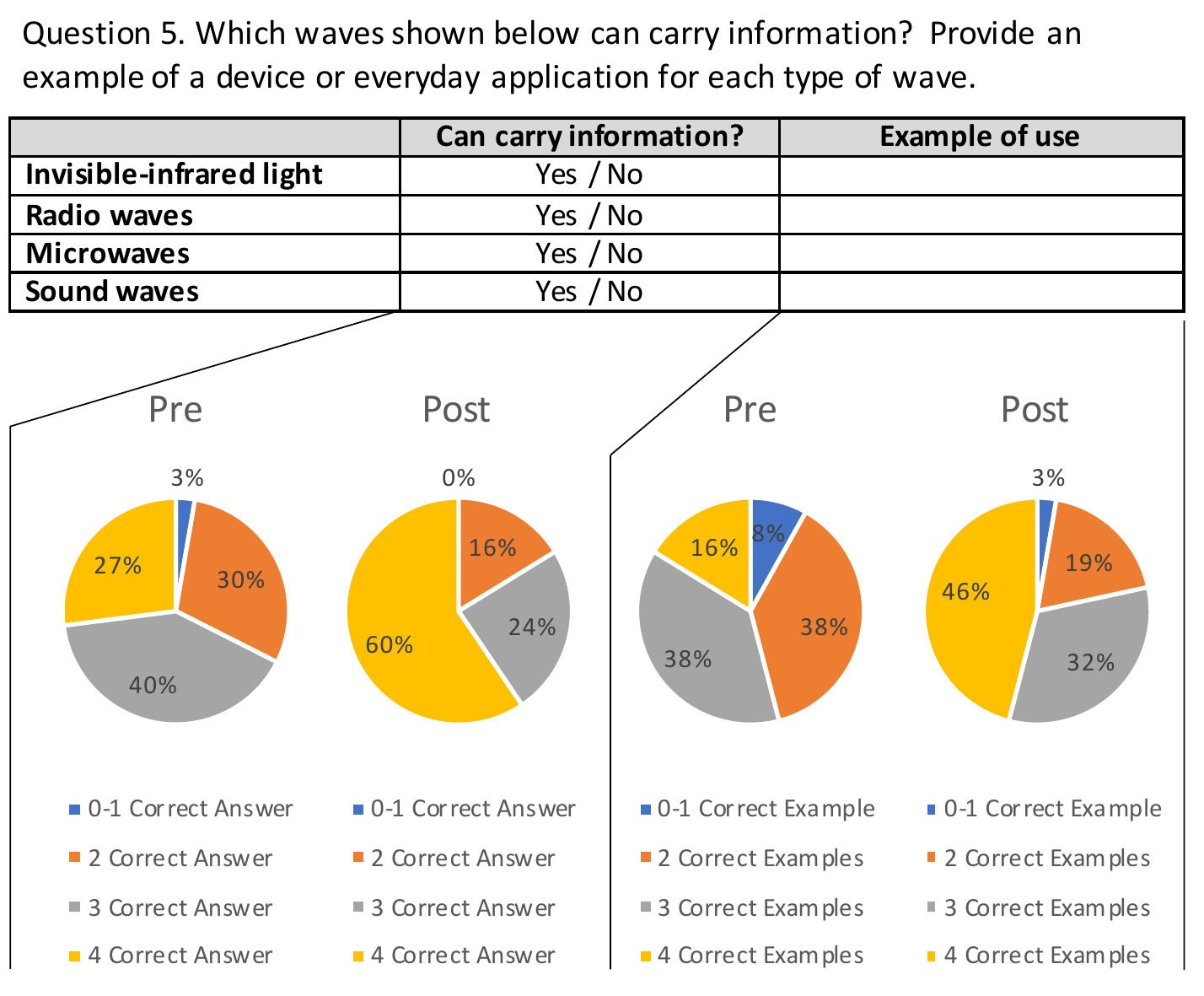}
\caption{Pre-post comparison of knowledge about wave transmission. Students identified which parts of the spectrum carry information and provided open-ended examples of each application.}
\label{fig:quest5}
\end{figure}

The pre-survey included five questions to assess student understanding of concepts related to 1) ordering the wavelengths of electromagnetic waves, 2) relating frequency and wavelength, 3) relating wavelength and energy, 4) identifying examples of transmission, reflection, and absorption phenomena, and 5) identifying which spectrum (i.e., infrared light, radio waves, microwaves, and sound waves) can carry information and providing examples of applications. The post-survey included the same five questions as well as overall experiences and future career questions. We report the results of student answers to questions one through four in Table \ref{tab:assign} and answers to question 5 in Fig. \ref{fig:quest5}.
\begin{table}[t]
    \centering
   \caption{Pre-post assessment of content knowledge. C and I stand for correct and incorrect answers, respectively.}
\includegraphics[width=0.44\textwidth]{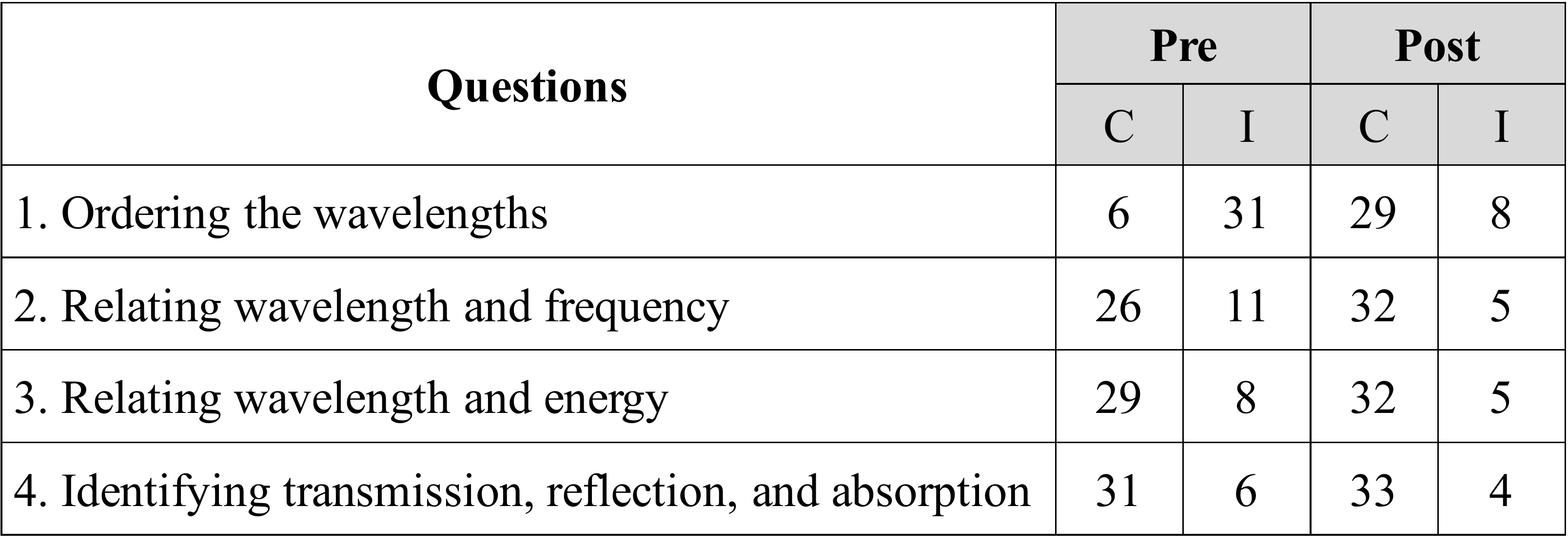}
    \label{tab:assign}
\end{table}

Table \ref{tab:assign} reports a comparison of students' pre-post responses to the content knowledge items for questions one through four. The pre-survey reveal that students were already familiar with the relationship between wavelength and energy, as well as able to identify examples of transmission, reflection, and absorption. However, most students (89\%) were unable to order the electromagnetic wave spectrum prior the instruction and there was substantial improvement following instruction.  Overall, most students were able to relate wavelength to frequency and energy prior to instruction; however, for a few students the activity appears to be a sufficient refresher.

Fig. \ref{fig:quest5} shows the distribution of scores when students were asked to identify which waves (i.e., infrared light, radio waves, microwaves, and sound wave) can carry information and provided examples of applications. Comparison of pre-post responses shows a pronounced change in student ability to identify and provide examples following instruction. Prior instruction, few students (27\%) recognized that all four waves carry information and even fewer students (16\%) were able to provide correct examples of each.  Following instruction, more students (60\%) correctly responded that all four waves carry information and more students (46\%) were able to provide examples of each application. Additional item analyses revealed that 15\% of students held misconceptions primarily about microwaves and sound waves following instruction. While students were familiar with microwaves from daily use of a common kitchen appliance, they were surprised during instruction to learn that Wi-Fi routers use the same frequency.

\subsection{Overall Experience with DIY-LWS Kit}
The post survey asked students to rate their overall experiences using the DIY-LWS kits on a 10-pt scale, including ease of use [M = 2.3, SD = 1.3, (1) Easiest – I was able without any help: (10) Hardest – I could not do any step] and comprehension [M = 8.4, SD = 1.3, (1) I was able to understand none of it: (10) I was able to understand it all]. Most students (81\%) agreed that the experience was enjoyable and shared their excitement through comments like, ``Wow! But how?'', ``Mind-blowing'', and ``This is black magic''. Additionally, students were asked to reflect about the influence of the experience on their evolving career interests in engineering in general [M = 4.7, SD = 0.6, (1) I am not interested in pursuing an engineering career: (5) I am strongly interested in pursuing a career in engineering], and sensing, communication, and optics technologies in particular [M = 3.7, SD = 0.9, (1) As an area of study, I am not interested in sensing, communication, or optics technologies: (5) As an area of study, I am strongly interested in sensing, communication, or optics technologies]. The vast majority of students (97\%) appreciated for being selected for the opportunity and most students (84\%) reported having more confidence in majoring in engineering as a result. 

\subsection{Discussion}
The DIY-LWS kit was successful in improving understanding about sensing and communication technologies. The DIY-LWS kit was successful in improving understanding about the frequencies and transmission capabilities of other wavelengths, such as microwaves, x-rays, and radio waves. Students found the kits easy to use and to understand, while enjoying their experience and raising their confidence in majoring in engineering. The kit is simple to assemble and is engaging with music.

Our future research and evaluation will identify and explore the reasons for misconceptions about the infrared light and microwave spectrum, which elicited mostly incorrect responses prior to the survey. Our assessment and evaluation demonstrated that the DIY-LWS is an exciting way to explore the basics of Li-Fi applications and their increasing relevance in the field of engineering.

We also note that our setup is safe, requiring low power batteries and dim LED lights. After the program, we allowed the students to take the DIY-LWS kit home and show their friends, teachers, and parents, for which we received very positive feedback from the students later. This also opens a possibility to be replicated by numerous elementary and high schools and universities around the world. 

\vspace{-2mm}
\section{Conclusion} \label{conclusion}
In this study, we have provided our experience with DIY-LWS kit with the required components and assemble instruction. We also reported the results of a survey applied prior to and after student participation. We have observed that a low-cost, effective, portable and easy to access DIY wireless sensing hands-on design project can help students learn the basic working principles behind EM sensing and communication systems.  
The kits further increased student interest and participation in related engineering courses. 
Our effort is particularly important to recruit engineering students in sensing and communication fields at the age of 5G and mmWave wireless communication and ever increasing mobile users and data transfer.   

% Moreover, overall students' response to the DIY-LWS kit was fantastic. ``Wow! But how?'', ``Mind-blowing'', and ``This is black magic. How does this work?'' are just three of the reactions that we received during the outreach events that is showing their enthusiasms. These comments describe how our effort triggered the curiosity of students and their teachers, which is a powerful motivation for future research and development. 

\vspace{-2mm}
\section*{Acknowledgment}
The authors would like to thank all their undergraduate, graduate students and staff members of the outreach programs at Oklahoma State University for their great support during the events that these DIY kits are used, such as National Lab Day, Discovery Days, and Summer Bridge programs. 
\vspace{-2mm}

\bibliographystyle{IEEEtran}

\bibliography{arXiv_ToE_DIY_Light-wave_Sensing_Comm}

%\newpage

%\footnotesize{\textbf{Sabit Ekin (M'12)}  received the B.Sc. degree in electrical and electronics engineering from Eski\c sehir Osmangazi University, Turkey, in 2006, the M.Sc. degree in electrical engineering from New Mexico Tech, Socorro, NM, USA, in 2008, and the Ph.D. degree in electrical and computer engineering from Texas A\&M University, College Station, TX, USA, in 2012. He was a Visiting Research Assistant with the Electrical and Computer Engineering Program, Texas A\&M University at Qatar from 2008 to 2009. In summer 2012, he was with the Femtocell Interference Management Team in the Corporate Research and Development, New Jersey Research Center, Qualcomm Inc. He joined the School of Electrical and Computer Engineering, Oklahoma State University, Stillwater, OK, USA, as an Assistant Professor, in 2016. He has four years of industrial experience from Qualcomm Inc., as a Senior Modem Systems Engineer with the Department of Qualcomm Mobile Computing. At Qualcomm Inc., he has received numerous Qualstar awards for his achievements/contributions on cellular modem receiver design. His research interests include the design and performance analysis of wireless communications systems in both theoretical and practical point of views, interference modeling, management and optimization in 5G, mmWave, HetNets, cognitive radio systems and applications, satellite communications, visible light sensing, communications and applications, RF channel modeling, non-contact health monitoring, and Internet of Things applications.}

\end{document}